\documentclass{JHEP3}

\title{Non-relativistic metrics with extremal limits}

\author{Emiliano Imeroni\\
Physique Th\'eorique et Math\'ematique, Universit\'e Libre de Bruxelles\\
\& International Solvay Institutes, CP 231, 1050 Bruxelles, Belgium\\
E-mail: \email{eimeroni@ulb.ac.be}}

\author{Aninda Sinha\\
Perimeter Institute for Theoretical Physics, Waterloo, Ontario N2L 2Y5, Canada\\
E-mail: \email{asinha@perimeterinstitute.ca}}

\abstract{We present solutions of type IIB supergravity with $z=2$ Schr\"odinger asymptotics that admit an extremal limit, $\ie$ the black hole horizon has a double zero. These solutions are obtained as TsT transformations of the charged planar black hole in $\text{AdS}_5 \times  S^5$. Unlike the uncharged solution, the Ramond-Ramond two-form is turned on. We study the thermodynamic properties of these new solutions, and we show that the ratio of shear viscosity to entropy density is $1/4\pi$ even in the extremal limit. We also consider the TsT-transformed soliton and show that, for a special radius of the compact circle, there is a confinement-deconfinement phase transition at zero temperature between the soliton and black hole phases.}


\preprint{}

\usepackage{amsmath,amssymb,amsfonts}
\usepackage{graphicx}

\newcommand{\ls}{\ensuremath{{l_s}}} 
\newcommand{\lp}{\ensuremath{{l_p}}} 
\newcommand{\gs}{\ensuremath{{g_s}}} 
\newcommand{\cp}{{\ensuremath{\mathbb{CP}^2}}} 
\newcommand{\hd}[1]{\ensuremath{\phantom{ }^{\star_{#1}}}} 

\newcommand{\hgamma}{{\hat{\gamma}}}

\newcommand{\M}{\mathcal{M}}

\def\half {\frac{1}{2}}
\def\be{\begin{equation} }
\def\ee{\end{equation}}
\def\ba{\begin{eqnarray} }
\def\ea{\end{eqnarray}}
\def\lp{\ell_P}
\def\ie{\it{i.e.}}

\begin{document}

\section{Introduction}\label{s:intro}

During the past year, considerable effort has been focussed on extending the AdS/CFT correspondence to examples where the associated symmetry algebra includes the Schr\"odinger algebra~\cite{son, bala}. The asymptotic metric in this case reads:%
\footnote{In the introduction we put for simplicity the AdS radius $l$ to $1$.}
\begin{equation}\label{Schr}
	ds^2 = r^2 \left( - 2 du dv -r^{2 z-2} du^2 + d\mathbf{x}^2\right) + \frac{dr^2}{r^2} + ds^2_M\,,
\end{equation}
where $ds_{M}^2$ is the metric of an appropriate compact manifold which allows~\eqref{Schr} to be a solution to the supergravity equations of motion. The direction $u$ is identified with the CFT time, while $v$ is taken to be compact so that the associated quantum number can be taken to be the particle number. As such, for non-relativistic setups, the duality is between a $d$ dimensional CFT and a $d+2$ dimensional gravity. The isometry algebra is generated by the following Killing vectors:
\begin{equation}
\begin{gathered}
	M_{ij}=-i( x^i \partial_j-x^j \partial_i)\,,\quad
	P_i=-i\partial_i\,,\quad
	H=-i \partial_u\,,\quad
	K_i=-i(-u \partial_i+x^i\partial_v)\,,\\
	D=-i (z u \partial_u +x^i \partial_i +(2-z)v\partial_v - r \partial_r)\,, \quad
	N=-i\partial_v\,.
\end{gathered}
\end{equation}
Here $z$ is called the dynamical exponent. The usual AdS case corresponds to $z=1$, with only the metric and the five-form flux being nontrivial and all other fields set to zero. Solutions with dynamical exponent $z = 2$ can be obtained starting from $z = 1$ by means of a TsT transformation~\cite{mmt,abm,hmr}. In addition to the metric and five-form, now one also has the NS-NS two-form turned on. In this case there is an additional Killing vector:
\be
	C=-i \left(u^2\partial_u+ u x^i \partial_i
		-\tfrac{r^2 {\mathbf{x}^2}+1}{2 r^2}\partial_v - r u \partial_r \right)\,,
\ee
which realizes the special conformal extension of the $z = 2$ algebra.

The extension to the case of finite temperature is obtained by starting with the AdS-Schwarzschild black hole~\cite{mmt,abm,hmr}. Here there is also a nontrivial dilaton in the background. The dispersion relations for excitations look like $\omega \sim k^2$ or, in other words, the particles are non-relativistic. There have been many further studies of related solutions \cite{lots}. $1/N$ corrections to $z$ were investigated in~\cite{amsv}.

Although the original motivation to realize a dual description of fermions at unitarity still defies an answer, some lessons have been learnt about solutions with asymptotics that describe non-relativistic systems. In some sense, we have enlarged the landscape of the AdS/CFT correspondence. It is of enormous relevance to try to find the universal features of this non-relativistic sector of the landscape, with the hope of being able one day to ask and answer the ``right" questions. In this spirit, this paper investigates one possible extension of the AdS/CFT correspondence for non-relativistic theories, that consists of turning on an extra chemical potential in the CFT, corresponding to an extra charge in the dual supergravity description.

Another motivation for pursuing this line of questioning is as follows. Recently, a proposal has been made to study Fermi surfaces using the AdS/CFT correspondence~\cite{liuetal}. In the example studied in~\cite{liuetal}, one considers a charged planar AdS black hole which admits an extremal limit, $\ie$ the horizon has a double zero when the temperature $T$ goes to zero. The Fermi surface is revealed by the appearance of gapless excitations of fermionic composite operators. The presence of a double zero appears to play a crucial role in this analysis. A question to ask is how to extend this construction for non-relativistic systems.%
\footnote{Fermions in non-relativistic systems have been studied in~\cite{fermions}.}

One obvious way to tackle this is to consider the charged planar AdS black hole and consider its TsT transformation, in the context of type IIB supergravity. Our starting point is the ten dimensional uplift of the supergravity solution considered in~\cite{mps09}, using the prescription in~\cite{myers}. The novelty in this case is that, in addition to the metric, five-form, NS-NS $B$-field and dilaton, the R-R two-form is also turned on. We present the derivation of this solution and study some of its properties.

We investigate the near horizon geometry of the TsT-transformed charged black hole by presenting two configurations of fields which satisfy the full ten-dimensional equations of motion in the region near the horizon. The first of these two solutions, which is obtained by means of a TsT transformation of the undeformed near horizon geometry, lacks the AdS$_2$ factor that is usually characteristic of a double zero horizon~\cite{hr1,hr2}. Its isometries only involve $P_i$, $H$ and $N$. The second solution, obtained via a scaling limit following the definition of near horizon limit given in~\cite{hr1,hr2}, has the AdS$_2$ isometries but can be shown to be also the near horizon limit of another black hole solution, without Schr\"odinger asymptotics, whose gauge dual is a standard non-commutative dipole theory.

We then consider the thermodynamics of the transformed charged black hole. We present an analysis in 10 dimensions for the counterterms which reproduces known results and extend it to our case. The form of the free energy which follows from the Euclidean action is inherited from the AdS case very much like the zero-charge case.

Another metric that solves the equations of motion is that of the soliton which we will call the Schr\"odinger soliton. To obtain this, one performs a double Wick rotation of the uncharged solution. The resulting geometry no longer has a horizon. Regularity of the Wick rotated time coordinate now gives us a periodic spatial coordinate. The resulting solution is nothing but the TsT cousin of the AdS soliton~\cite{sol2}. If one imposes antiperiodic boundary conditions for fermions in the periodic direction, one can look for a phase transition between the uncharged black hole and the soliton. This was studied in~\cite{Mann:2009yx}, where it was claimed that the black hole is the preferred phase at high temperatures, while the soliton is preferred at low temperatures. We re-examine this case and find that, due to a relation between the non-relativistic deformation parameter and the parameters of the soliton solution, a subtlety with the boundary conditions prevents any phase transition from occurring except for a special radius of the periodic spatial coordinate. We extend this analysis to the charged case. We show that there is a confinement-deconfinement transition at zero temperature between the black hole and soliton phases at the special radius.

The paper is organized as follows. We present the TsT-transformed charged black hole solution in section~\ref{s:solution}, then we consider its near horizon geometry in section~\ref{s:nh}. In section~\ref{s:thermo}, we turn to the thermodynamics of the black hole. We compute the ratio of shear viscosity to entropy density in section~\ref{s:visco}. In section~\ref{s:soliton}, we turn to the TsT-transformed soliton and study the phase structure of the theory. We conclude with a discussion in section~\ref{s:concl}. Three appendices complement the main text, appendix~\ref{s:conventions} outlining our conventions, appendix~\ref{s:nheoms} clarifying some aspects of near horizon solutions and finally appendix~\ref{s:isometries} studying the isometries of the configurations of section~\ref{s:nh}.

\vskip 0.5em
While this work was in progress, we became aware of the overlapping work by Adams et al.~\cite{adamsnew}.

\section{TsT of the charged AdS black hole}\label{s:solution}

We start with the type IIB solution describing a charged AdS black hole, which is the ten-dimensional uplift of the solution considered in~\cite{mps09}. The non-zero fields are the metric and the five-form, whose expressions read:
\begin{equation}\label{BH}
\begin{split}
	(ds^2)^{(0)} &= \frac{r^2}{l^2} \left( - f dt^2 + dx^2 + dy^2 + dz^2 \right)
		+ \frac{l^2}{r^2} \frac{dr^2}{f}\\
		& + l^2 \left( d\alpha^2 + \sin^2\alpha d\beta^2
		+ \mu_1^2 (d \xi_1 + A)^2 + \mu_2^2 (d \xi_2 + A)^2 + \mu_3^2 (d \xi_3 + A)^2 \right)\,,\\
	F_5^{(0)} &= (1 + \hd{}) \left[ \left( -\frac{4 r^3}{l^4} dt \wedge dr
		+\frac{Q}{l^2}\, d \left( \sum_{i=1}^3 \mu_i^2 d \xi_i \right)
		 \right) \wedge dx \wedge dy \wedge dz \right] \,,
\end{split}
\end{equation}
where
\begin{equation}
	\mu_1 = \cos \alpha\,,\qquad
	\mu_2 = \sin \alpha \cos \beta\,,\qquad
	\mu_3 = \sin \alpha \sin \beta\,,
\end{equation}
and
\begin{equation}
	f(r) = \left( 1 - \frac{r_0^2}{r^2} \right)
		\left( 1+ \frac{r_0^2}{r^2} - \frac{Q^2}{r_0^2 r^4} \right)\,,\qquad
	A = A_t\, dt = \frac{Q}{l^2} \left( \frac{1}{r_0^2} - \frac{1}{r^2} \right) dt \,.
\end{equation}

Starting from~\eqref{BH}, we can obtain a solution with non-relativistic Schr\"odinger asymptotics~\eqref{Schr} by performing a ``TsT transformation''~\cite{mmt} (also called a ``null Melvin twist''~\cite{abm,hmr}) along the two-torus parameterized by the light-cone direction $x^- = \frac{1}{2} \left( t - x \right)$ and by an internal angle $\psi$. The TsT transformation~\cite{Lunin:2005jy} consists of a T-duality along $\psi$, followed by a shift $x^- \to x^- - \gamma \psi$ and finally by another T-duality along $\psi$.

General explicit formulae for the TsT transformation of any type II background were given for instance in~\cite{ei} and are reviewed in appendix~\ref{s:conventions}. In the case at hand, where we start from a solution with just the metric and $F_5$ turned on, we do not need the TsT formulae in full generality, but just the particular case in which we are able to put the undeformed metric in the form:
\begin{equation}\label{easyform}
	(ds^2)^{(0)} = ( A dx^- + K_1)^2 + (B d \psi + C dx^- + K_2)^2 + ds^2_8\,,
\end{equation}
where $ds^2_8$ and the one-forms $K_1$, $K_2$ do not depend on $x^-$ and $\psi$. If this is the case, the NS-NS part of the TsT-transformed solution in the string frame is given by:
\begin{equation}\label{NSNSTsT}
\begin{split}
	ds^2 &= \M ( A dx^- + K_1)^2 + \M (B d \psi + C dx^- + K_2)^2 + ds^2_8\,,\\
	e^{2 \Phi} &= \M\,,\\
	B &= - \gamma \M\, A B\, ( A dx^- + K_1) \wedge (B d \psi + C dx^- + K_2)\,,
\end{split}
\end{equation}
where $\M = (1 + \gamma^2 A^2 B^2)^{-1}$. Next, the non-vanishing R-R field strengths $\mathcal{F}_p = F_p + H \wedge C_{p-3}$ of the transformed solution, in the case where the starting solution has only the five-form turned on, are given by:
\begin{equation}\label{RRTsT}
	{F}_3 = \gamma\, \iota_{x^-}\, \iota_{\psi}\, {F_5}^{(0)}\,,\qquad
	\mathcal{F}_5 + {F}_3 \wedge {B}
		=  {F_5}^{(0)}\,,\qquad
\end{equation}
where $\iota$ denotes the interior product.

Let us then put the solution~\eqref{BH} in the form~\eqref{easyform}. Towards this goal, we first define $x^\pm = \frac{1}{2} \left( t \pm x \right)$ and introduce the following vielbeins for the AdS part:
\begin{equation}\label{eAdS}
\begin{split}
	e^0 &= \frac{r}{l} \sqrt{\frac{4 f}{1-f}} dx^+\,,\\
	e^1 &= \frac{r}{l} \frac{1}{\sqrt{1-f}} \left( (1-f) dx^- - (1+f) dx^+ \right)\,,\\
	e^2 &= \frac{r}{l} dy\,,\qquad e^3 = \frac{r}{l} dz\,,\qquad e^4 = \frac{l}{r} \frac{dr}{\sqrt{f}}\,,
\end{split}
\end{equation}
and for the sphere part:
\begin{equation}\label{eS5}
\begin{split}
	e^5 &= l d\alpha\,,\qquad e^6 = l \sin \alpha\, d \beta\,,\\
	e^7 &= l \sin \alpha \cos \alpha
		\left( \cos^2 \beta\, (d\xi_1 - d\xi_2) + \sin^2 \beta\, (d\xi_1 - d\xi_3 ) \right)  \,,\\
	e^8 &= l \sin \alpha \sin \beta \cos \beta\, (d\xi_2 - d\xi_3) \,,\\
	e^9 &= l \left(\cos^2 \alpha\, d\xi_1 + \sin^2 \alpha \cos^2 \beta\, d\xi_2
		+ \sin^2 \alpha \sin^2 \beta\, d\xi_3 + A \right)\,.
\end{split}
\end{equation}
Notice that the vielbeins~\eqref{eS5} highlight the structure of the metric as a fiber bundle over \cp{} (see for example \cite{sss}). In fact, if we define a new angle $\psi = \frac{1}{3} (\xi_1 + \xi_2 + \xi_3)$, we can write $e^9 = l (d\psi + P + A)$ and $ds^2_{\cp} = (e^5)^2 + (e^6)^2 + (e^7)^2 + (e^8)^2$ so that the sphere part of the metric becomes:
\begin{equation}
	(e^5)^2 + \ldots + (e^9)^2 = l^2 (d\psi + P + A)^2 + ds^2_{\cp} \,.
\end{equation}

The solution~\eqref{BH} in terms of the vielbeins we have defined reads:
\begin{equation}\label{BHe}
\begin{split}
	(ds^2)^{(0)} &= \eta_{ab} e^a e^b \,,\\
	F_5^{(0)} &= (1 + \hd{}) \Big[ -\frac{4}{l} e^0\, \wedge e^1 \wedge e^2 \wedge e^3 \wedge e^4\\
		&\qquad +\frac{2 Q}{l r^3 \sqrt{1-f}}\, (\sqrt{f} e^0 + e^1)
		\wedge e^2 \wedge e^3 \wedge \omega_{\cp} \Big] \,,
\end{split}
\end{equation}
where $\omega_{\cp} = -\tfrac{l^2}{2} dP = e^5 \wedge e^7 + e^6 \wedge e^8$ is the K\"ahler form on \cp.

We are now ready to apply the TsT transformation along $(x^-,\psi)$ with real parameter $\gamma$, by applying the formulae~\eqref{NSNSTsT}-\eqref{RRTsT} to the solution in the form~\eqref{BHe}. The TsT-transformed background, in the string frame and in terms of the undeformed vielbeins, reads:
\begin{equation}\label{TsTe}
\begin{aligned}
	ds^2 &= - (e^0)^2 + \M (e^1)^2 + (e^2)^2 + \ldots +  (e^8)^2 + \M  (e^9)^2\,,\\
	e^{2 \Phi} &= \M = \left( 1 + \gamma^2 r^2 (1-f) \right)^{-1}\,,\\
	B &= \gamma r \sqrt{1-f} \M e^1 \wedge e^9\,,\\
	C_2 &= - \gamma l^2 A_t \, \omega_{\cp}\,, \qquad
	F_3 = -\frac{2 \gamma Q \sqrt{f}}{l r^2} e^4 \wedge \omega_{\cp}\,,\\
	\mathcal{F}_5 &= F_5 + H_3 \wedge C_2 = (1 + \hd{})\, G_5\\
		& = (1 + \hd{}) \Big[ -\frac{4}{l}\, e^0 \wedge e^1 \wedge e^2 \wedge e^3 \wedge e^4
		+\frac{2 Q}{l r^3 \sqrt{1-f}}\, (\sqrt{f} e^0 + e^1)
		\wedge e^2 \wedge e^3 \wedge \omega_{\cp} \Big] \,,
\end{aligned}
\end{equation}
Notice that, although the expression for $\mathcal{F}_5$ is the same as the one for $F_5$ in the undeformed background, the Hodge $\hd{}$ is now performed with the deformed metric in~\eqref{TsTe}. The result is that the $e^1\wedge e^9$ components of the five-form are multiplied by a factor $\M$ with respect to the undeformed case. The solution~\eqref{TsTe} can be explicitly checked to satisfy the type IIB equations of motion in our conventions written in appendix~\ref{s:conventions}.

It will also be useful to compute the four-form potential $C_4$, given by:
\begin{equation}\label{C4}
\begin{split}
	C_4 &= \frac{1}{\sqrt{f}} \left( -1 + \frac{Q A_t l^2}{r^4} \right)
		e^0 \wedge e^1 \wedge e^2 \wedge e^3
		+ \frac{Q}{r^3 \sqrt{1-f}} \left( \sqrt{f} e^0 + e^1 \right)
		\wedge e^2 \wedge e^3 \wedge e^9\\
		&\qquad + \tan \alpha\, e^6 \wedge e^7 \wedge e^8 \wedge
		\left( e^9 + \frac{Q}{r^3 \sqrt{1-f}} \left( \frac{1}{\sqrt{f}} e^0 + e^1 \right) \right)\\
		&\qquad +  \frac{Q}{r^3 \sqrt{1-f}} \left( \frac{1}{\sqrt{f}} e^0 + \M e^1 \right)
		\wedge e^9 \wedge \left( e^5 \wedge e^7 + e^6 \wedge e^8 \right)\,.
\end{split}
\end{equation}
The above result is obtained starting from the undeformed potential $C_4^{(0)}$ of the solution~\eqref{BH} via $C_4 + C_2 \wedge B = C_4^{(0)}$, see~\eqref{RRpotTsT}. The expression of $C_4^{(0)}$ is given by the same expression~\eqref{C4} without the factor of $\M$ in the last line, since we have:
\begin{equation}
	B \wedge C_2 = - (\M - 1) \frac{Q}{r^3 \sqrt{1-f}} e^1
		\wedge e^9 \wedge \left( e^5 \wedge e^7 + e^6 \wedge e^8 \right)\,.
\end{equation}

Going back to the original coordinates of the undeformed background~\eqref{BH}, we can rewrite our TsT-transformed charged black hole~\eqref{TsTe} in the string frame as follows:
\begin{equation}\label{TsT}
\begin{split}
	ds^2 &=\frac{r^2}{l^2} \M
		\left( - f dt^2 + dx^2 - \gamma^2 r^2 f (dt + dx)^2 \right)
		+ \frac{r^2}{l^2} \left(dy^2 + dz^2 \right) + \frac{l^2}{r^2} \frac{dr^2}{f}\\
		&\qquad\qquad + l^2 \left( \M\, (d\psi + P + A)^2 + ds^2_{\cp} \right)\,,\\
	e^{2 \Phi} &= \M\,,\\
	B &= - \gamma r^2 \M\, (f dt + dx) \wedge (d\psi + P + A)\,,\\
	C_2 &= - \gamma l^2 A_t \, \omega_{\cp}\,, \\
	\mathcal{F}_5 &= (1 + \hd{})\, G_5 = (1 + \hd{}) \left[ \left( -\frac{4 r^3}{l^4} dt \wedge dr
		- \frac{2Q}{l^4}\, \omega_{\cp} \right) \wedge dx \wedge dy \wedge dz \right] \,,
\end{split}
\end{equation}
where we recall the definition of the function $\M$:
\begin{equation}
	\M = \left( 1 + \gamma^2 r^2 (1-f) \right)^{-1}\,.
\end{equation}
The solution~\eqref{TsT} reduces to the known uncharged non-relativistic solution when $Q=0$ and (obviously) to the charged black hole~\eqref{BH} when $\gamma=0$. Introducing the coordinate system
\begin{equation}\label{uv}
	u = 2 \gamma l\, x^+ = \gamma l \left(t + x\right)\,,\qquad
	v = \frac{1}{\gamma l}\, x^+ = \frac{1}{2 \gamma l} \left(t - x\right),
\end{equation}
which eliminates the parameter $\gamma$ from the asymptotic expression at $r \to \infty$,  one recovers the Schr\"odinger metric~\eqref{Schr} in the limit. The coordinate $u$ in~\eqref{uv} is interpreted as the CFT time, while $v$ is taken to be periodic to make the number operator $N$ have discrete eigenvalues.

\section{Near horizon geometry}\label{s:nh}

As noted in the introduction, one of the main reasons for us to study the solution presented in section~\ref{s:solution} is the possibility of having a double zero at the horizon and hence an extremal limit corresponding to zero temperature. This happens when $Q=\sqrt{2} r_0^3$. In this section we consider the near horizon geometry of our solution in this extremal limit.

Let us first take a step back and start from the charged AdS black hole~\eqref{BH} before the TsT transformation, namely from the relativistic setup. We write $r = r_0 + \rho$ and expand the fields of the solution, obtaining the following near horizon configuration that solves the equations of motion of type IIB supergravity:
\begin{equation}\label{nhbh}
\begin{split}
	(ds^2)^{(0)} &=\tfrac{r_0^2}{l^2}
		\left( - \tfrac{12 \rho^2}{r_0^2} dt^2 + dx^2 + dy^2 + dz^2 \right)
		+ \tfrac{l^2}{12\rho^2} d\rho^2\\
		&\qquad + l^2 \left( \left(d\psi + P + \tfrac{2\sqrt{2}\rho}{l^2} dt \right)^2
		+ ds^2_{\cp} \right)\,,\\
	F_5^{(0)} &= (1 + \hd{}) \left[ \left( -\tfrac{4 r_0^3}{l^4} dt \wedge d\rho
		- \tfrac{2\sqrt{2}r_0^3}{l^4}\, \omega_{\cp} \right) \wedge dx \wedge dy \wedge dz \right] \,.
\end{split}
\end{equation}
The $(t,\rho)$ coordinates clearly span an AdS$_2$ space, which signals an enhancement of the isometries of the near horizon geometry to $SO(2,1)$. This is a peculiar feature of the near horizon region of extremal black holes~\cite{hr1,hr2}. The presence of this AdS$_2$ factor has been crucial for a proposal to study Fermi surfaces using the AdS/CFT correspondence~\cite{liuetal}, and we want to study what happens in the non-relativistic case of section~\ref{s:solution}.

Notice that expanding a black hole solution near its horizon, $r = r_0 + \rho$, is not guaranteed to yield a solution to the supergravity equations of motion. In particular, we are unable to find a suitable truncation of the non-extremal solution of section~\ref{s:solution}, or of the usual AdS-Schwarzschild case. However, it has been shown in the relativistic setup that any extremal black hole admits a near horizon limit~\cite{hr1,hr2}, a fact that we try to explain in more detail in appendix~\ref{s:nheoms}.

In the case of the extremal limit of~\eqref{TsT}, a naive expansion of the metric and the potentials, by means of which we would for instance put  $\M \to \left( 1 + \gamma^2 r_0^2\right)^{-1}$ directly, does not seem to yield a solution to the type IIB equations. In order to find the near horizon geometry, we then follow a more indirect route that will also allow us to uncover interesting properties of two different solutions along the way. We start from the near horizon limit~\eqref{nhbh} of the undeformed background and apply the TsT transformation along $\left(x ^- = \frac{1}{2} \left( t - x \right) ,\psi \right)$, in complete analogy to the procedure followed in the previous section. After the transformation we obtain the following configuration in the string frame:
\begin{equation}\label{TsTnh}
\begin{split}
	ds^2 &=\tfrac{r_0^2}{l^2} \M_0
		\left( - \tfrac{12 \rho^2}{r_0^2} dt^2 + dx^2 - 12\gamma^2 {\rho^2}   (dt + dx)^2 \right)
		+ \tfrac{r_0^2}{l^2} \left(dy^2 + dz^2 \right) + \tfrac{l^2}{12 \rho^2} d\rho^2 \\
		&\qquad\qquad + l^2 \left( \M_0\, \left(d\psi + P + \tfrac{2\sqrt{2}\rho}{l^2} dt \right)^2
		+ ds^2_{\cp} \right)\,,\\
	e^{2 \Phi} &= \M_0\,,\\
	B &= - \gamma r_0^2 \M_0\, \left( \tfrac{12 \rho^2}{r_0^2} dt + dx \right) \wedge
		\left(d\psi + P + \tfrac{2\sqrt{2}\rho}{l^2} dt \right)\,,\\
	C_2 &= - 2 \sqrt{2} \gamma \rho\, \omega_{\cp}\,, \\
	\mathcal{F}_5 &= (1 + \hd{}) \left[ \left( -\tfrac{4 r_0^3}{l^4} dt \wedge d\rho
		- \tfrac{2\sqrt{2}r_0^3}{l^4}\, \omega_{\cp} \right) \wedge dx \wedge dy \wedge dz \right] \,,
\end{split}
\end{equation}
where
\begin{equation}\label{corrM0}
	\M_0 = \left( 1 + \gamma^2 (r_0^2 - 12 \rho^2)\right)^{-1}\,.
\end{equation}
The above geometry does (obviously) solve the IIB equations of motion. We also see that it is in fact an expansion of the full solution~\eqref{TsT} around the horizon $r = r_0$, although maybe not the most naive one could have imagined.

What has happened to the isometries of the undeformed near horizon solution~\eqref{nhbh} after the TsT transformation? As usual, the transformation has reduced the amount of symmetry of the background down to the symmetries that ``commute'' with it~\cite{Lunin:2005jy}, so the AdS$_2$ factor of~\eqref{nhbh} has disappeared, together with the full $SO(2,1)$ isometry. It can be shown after a bit of work that the most general coordinate transformation that leaves the $x$, $t$, $\rho$ part of the metric invariant is the trivial translation of $x$ and $t$. This is demonstrated explicitly in appendix~\ref{s:isometries}. This is the parallel phenomenon, now happening near the horizon of the black hole, to the one happening in the full solution, where the asymptotic Poincar\'e and conformal symmetry of~\eqref{BH} are replaced by the non-relativistic asymptotic Schr\"odinger symmetry of~\eqref{TsT}. In this sense, then, we can say that~\eqref{TsTnh} represents the ``non-relativistic near horizon limit'' of our extremal black hole.

However, there is a way to obtain a near horizon geometry that keeps the AdS$_2$ factor following the definition of near horizon limit given in~\cite{hr1,hr2}.%
\footnote{We thank Allan Adams for discussions on this issue.}
This definition involves a specific scaling limit. Starting from~\eqref{TsTnh}, we rescale $t\to t/\lambda$, $\rho\to \lambda \rho$ and we consider the $\lambda\to 0$ limit. This leads to the following solution:
\begin{equation}\label{scale}
\begin{split}
	ds^2 &=\tfrac{r_0^2}{l^2}
		\left( - \tfrac{12 \rho^2}{r_0^2} dt^2 + \M_x\, dx^2 \right)
		+ \tfrac{r_0^2}{l^2} \left(dy^2 + dz^2 \right) + \tfrac{l^2}{12 \rho^2} d\rho^2\\
		&\qquad\qquad + l^2 \left( \M_x\, \left(d\psi + P + \tfrac{2\sqrt{2}\rho}{l^2} dt \right)^2
		+ ds^2_{\cp} \right)\,,\\
	e^{2 \Phi} &= \M_x\,,\\
	B &= - \gamma r_0^2\, \M_x\, dx \wedge
		\left(d\psi + P + \tfrac{2\sqrt{2}\rho}{l^2} dt \right)\,,\qquad C_2 = 0\,, \\
	F_5 &= (1 + \hd{}) \left[ \left( -\tfrac{4 r_0^3}{l^4} dt \wedge d\rho
		- \tfrac{2\sqrt{2}r_0^3}{l^4}\, \omega_{\cp} \right) \wedge dx \wedge dy \wedge dz \right] \,,
\end{split}
\end{equation}
where the dilaton is now constant, since we have:
\begin{equation}\label{M02}
	\M_x = \left( 1 + \gamma^2 r_0^2\right)^{-1}\,.
\end{equation}
This solution is not equivalent to~\eqref{TsTnh}, having been obtained via a scaling limit as described above.

We can see that we have recovered the AdS$_2$ factor in the geometry.  While \eqref{scale} is a solution of the equations of motion and an expansion of our extremal black hole solution (from which it could have been obtained directly by means of an appropriately defined scaling limit), it has a property that makes us wonder about its meaning in the non-relativistic context. This property is that the scaled solution~\eqref{scale} is also simply a TsT transformation of the near horizon geometry~\eqref{nhbh} of the undeformed charged black hole, but performed along the torus $(x, \psi)$ rather than $(x^-, \psi)$, as can be easily seen by using formulae~\eqref{NSNSTsT}-\eqref{RRTsT}. In this sense, \eqref{scale} is also the near horizon limit of another black hole, that is the one obtained from~\eqref{BH} via a TsT transformation along $(x, \psi)$. The latter describes a different dual gauge theory at finite temperature, namely a non-commutative dipole theory of the kind described in~\cite{dipole}, which however does not have non-relativistic Schr\"odinger asymptotics, given that the transformation used to get to it does not involve the asymptotic light-cone direction $x^-$.

One possible interpretation of this fact is that there is a sector where the CFT dual to the non-relativistic extremal black hole and the dipole field theory share the same description, this being probably the $H \to 0$ sector corresponding to the scaling limit we implemented to get to~\eqref{scale}. Since the AdS$_2$ decouples from the asymptotic geometry, we expect to be able to describe these low energy states of both these field theories using the superconformal quantum mechanics dual to AdS$_2$.

The geometry~\eqref{TsTnh} seems instead to be more intimately related to the non-relativistic nature of the solution, and could be used for instance to describe more general states. It would be interesting to investigate if~\eqref{TsTnh} decouples in the same sense as AdS$_2$ and if the higher energy states can be described using this geometry. Since the proper distance from any point $r=R$ to the horizon is given by $\int_{r_0}^{R} dr/\sqrt{g_{rr}}$ and $g_{rr}$ has a double pole in the extremal limit at $r=r_0$, this will work out to be infinity. Hence, it is plausible that~\eqref{TsTnh} does describe a decoupled subsector of the CFT. We leave the detailed examination of these issues for future work.

\section{Thermodynamics}\label{s:thermo}

This section is devoted to the discussion of the thermodynamics of our solution. We Wick-rotate into Euclidean signature $t \to i\tau$. The correct identification of thermodynamic quantities require some care because of the identification of the coordinate $u$ in~\eqref{uv} as the CFT time, see for instance~\cite{hmr}. The Killing vector along the horizon is taken to have coefficient one in front of the $\partial/\partial u$ term:
\begin{equation}\label{tuv}
	\frac{\partial}{\partial u} + \frac{1}{2 \gamma^2 l^2} \frac{\partial}{\partial v}
		= \frac{1}{\gamma l} \frac{\partial}{\partial t} \,.
\end{equation}
so that the temperature has an extra factor of $\gamma l$ in the denominator with respect to what one would have expected from the periodicity of $\tau$:
\begin{equation}\label{T}
	T=\frac{r_0}{\pi \gamma l^3} \left(1-\frac{Q^2}{2 r_0^6} \right)\,.
\end{equation}
Note that $T=0$ in the extremal limit $Q=\sqrt{2}r_0^3$. Equation~\eqref{tuv} also allows us to derive the expression for the chemical potential $\mu_1$ associated with $\gamma$ (that we divide by a factor of $l$ so that it has the appropriate dimensions of energy):
\begin{equation}
	\mu_1=\frac{1}{2\gamma^2 l^3}\,.
\end{equation}
We have a second chemical potential $\mu_2$ associated with the charge $Q$. In order to define it, notice that the expression $A_t$ appearing in the undeformed solution~\eqref{BH} comes from the uplift of a five-dimensional gauge field~\cite{mps09}. In such five dimensional setting, it is then natural to identify the chemical potential with the $r\to\infty$ limit of $A_t$. In our case, although we did not present an explicit five-dimensional description, we keep the same prescription, the only difference being that we again take into account the identification of $u$ as the CFT time. Rewriting $A = A_t dt = A_u du + A_v dv$, we then define the chemical potential $\mu_2$ as:%
\footnote{Note that in order to match with the conventions of \cite{mps09}, we need to choose $L_*=\sqrt{3} L$ in (2.23) of that paper.}
\begin{equation}
	\mu_2= \lim_{r\rightarrow \infty} A_{u}= \frac{Q}{2 \gamma l^3 r_0^2}\,.
\end{equation}

Let us pass to the computation of the free energy, for which we need the on-shell action. We are going to work in the full ten dimensional space. With the field configuration~\eqref{TsT}, the Euclidean Einstein frame action we are going to use is (see appendix~\ref{s:conventions}):
\begin{equation}\label{action}
	I = - \frac{1}{2\kappa^2} \bigg[ \int d^{10}x \sqrt{g}\ R
		- \frac{1}{2} \int \Big( d\Phi \wedge \hd{} d\Phi
		+ e^{-\Phi}  H \wedge \hd{}H
		+ e^{\Phi} F_3 \wedge \hd{} F_3
		+ G_5 \wedge \hd{} G_5\Big) \bigg] \,,
\end{equation}
where we recall that $G_5$ is the five-form such that $\mathcal{F}_5 = (1 + \hd{}) G_5$. This term for the five-form flux, giving a non-zero contribution, has been taken into account following \cite{gkt}.

The on-shell action is divergent so we need to add counterterms at the boundary:
\begin{multline}\label{ct}
	I_{ct} = -\frac{1}{2 \kappa^2} \int d^{9}\xi\, \sqrt{h}
		\bigg(2K  + \delta_0+\delta_1 \Phi+\delta_2 \Phi^2
		+ \frac{\delta_3}{2} B_{ij} B^{ij}\\
		+ \frac{\delta_4}{2} {C_{(2)\, ij}} C_{(2)}^{ij}
		+ \frac{\delta_5}{24} {C_{(4)\, ijkl }} C_{(4)}^{ijkl} \bigg) \,.
\end{multline}
Here $\xi$ denotes the boundary coordinates, $h$ is the induced metric on the 9-dimensional space and $i,j,k,l$ denote all indices except the radial one. $K$ is the extrinsic curvature:
\begin{equation}
	K = \nabla_{M} N^{M} = \frac{1}{\sqrt{g}} \partial_{M} (\sqrt{g} N^{M})\,,
\end{equation}
with $N^{M}$ the outward normal to the boundary.

One could take the approach of \cite{abm, hmr} and consider a variational problem which fixes some of the $\delta_i$'s above. Here we will fix the coefficients indirectly, by demanding that, besides of course cancelling the divergences, they reproduce known results in specific limits. More precisely, we demand that known results for $\gamma \to 0$ (see for example \cite{mps09}) and for $Q \to 0$~\cite{abm,hmr} are reproduced. This reasoning fixes $\delta_0 = - \tfrac{6}{l}$, $\delta_1 - 2 \delta_3 = -\tfrac{3}{2l}$ (for instance we can choose $\delta_1 = -\tfrac{3}{2l}$ and $\delta_3=0$), $\delta_2 = -\tfrac{3}{16l}$ and $\delta_5 = 0$. When the dust settles, the total Euclidean action turns out to be:
\begin{equation}\label{IE}
	I_{\text{E}} = - \frac{\Delta \tau \Delta x V r_0^4}{2 \kappa^2}
		\left[1+\frac{Q^2}{r_0^6} \left(1+2\delta_4 \gamma^2 l r_0^2 \right) \right]\,.
\end{equation}
Here $\Delta \tau$ and $\Delta x$ are the periodicities of $\tau$ and $x$, while $V = V_2 V_5$ is the product of the infinite volume $V_2 = \Delta y \Delta z$ of the two spatial gauge theory directions and of the (dimensionless) volume $V_5$ of the internal compact space. Notice that the periodicity $\Delta v$ of the $v$ coordinate in~\eqref{uv} is related to $\Delta x$ for any fixed value of $u$ as $\Delta x = \gamma l\, \Delta v$.

Assuming we are in a grand canonical ensemble, we can define the grand potential starting from~\eqref{IE} as:
\begin{equation}
	W = I_{\text{E}}\, T = - \frac{\Delta v V r_0^4}{2 \kappa^2}
		\left[1+\frac{Q^2}{r_0^6} \left(1+2\delta_4 \gamma^2 l r_0^2 \right) \right]\,.
\end{equation}
The identification of the entropy computed from the grand potential with the Bekenstein-Hawking entropy of the black hole will allow us to fix the arbitrary coefficient $\delta_4$. In terms of the grand potential we have:
\begin{equation}\label{S1}
	S = - \frac{\partial W}{\partial T} \bigg\rvert_{\mu_1\,,\, \mu_2}\,.
\end{equation}
On the other hand, the area of the horizon in the metric~\eqref{TsT}, which is $\gamma$-independent, is given by $A_H = l^2 r_0^3 \Delta x V$, so that the entropy computed via the Bekenstein-Hawking formula is:
\begin{equation}\label{S2}
	S = \frac{\Delta x V\, 2 \pi l^2 r_0^3}{\kappa^2}\,.
\end{equation}
Matching~\eqref{S1} and~\eqref{S2} leads to $\delta_4 = 0$. Hence the final expression for the grand potential, which yields the entropy~\eqref{S2}, is:
\begin{equation}\label{grandW}
	W = - \frac{\Delta v V r_0^4}{2 \kappa^2}
		\left( 1+\frac{Q^2}{r_0^6} \right)\,.
\end{equation}
This is the same expression as in the undeformed charged AdS black hole case, as was already noted for the $Q=0$ case in~\cite{hmr, abm, Yamada:2008if}.

We can reexpress the grand potential in terms of the temperature and chemical potentials by substituting (see also~\cite{mps09}):
\begin{equation}\label{inverse}
	r_0 = \frac{\pi l^{3/2} T}{2 \sqrt{2 \mu_1}}
		\left( 1+ \sqrt{1+\frac{8 \mu_2^2}{\pi^2 T^2}}\right)\,,\ \
	\gamma = \frac{1}{\sqrt{2 \mu_1 l^3} }\,,\ \
	\frac{Q}{r_0^3} = \frac{4 \mu_2}{\pi T}
		\left( 1+ \sqrt{1+\frac{8 \mu_2^2}{\pi^2 T^2}}\right)^{-1}\,,
\end{equation}
and we can compute, besides the entropy, also the charges associated to $\mu_1$ and $\mu_2$, whose expressions are much more easily rewritten in terms of $r_0$, $\gamma$ and $Q$:
\begin{align}
	P_1& = - \frac{\partial W}{\partial \mu_1} \bigg\rvert_{T\,,\, \mu_2}
		= - \frac{\Delta x V\, 2 \gamma l^2 r_0^4 \left(1+\frac{Q^2}{r_0^6} \right)}{\kappa^2} \,,\\
	P_2 &= - \frac{\partial W}{\partial \mu_2} \bigg\rvert_{T\,,\, \mu_1}
		= \frac{\Delta x V\, 6 l^2 Q}{\kappa^2} \,.
\end{align}
The energy is given by:
\begin{equation}\label{energy}
	E = W + T S + \mu_i P_i = 	\frac{\Delta v V r_0^4}{2 \kappa^2}
		\left( 1+\frac{Q^2}{r_0^6} \right)\,,
\end{equation}
and since in the grand canonical ensemble the pressure is given in terms of the grand potential by $PV = -W$ we see from~\eqref{grandW} and~\eqref{energy} that the system satisfies the equation of state
\begin{equation}
	P V = E
\end{equation}
appropriate for a non-relativistic CFT in two spatial dimensions.

\section{$\eta/s$}\label{s:visco}

In this section we turn to the computation of the ratio of the shear viscosity to entropy density. The argument given in \cite{Iqbal:2008by} suggests that the ratio should again be $1/4\pi$. We verify that this is indeed the case.  We begin by computing the effective action arising from~\eqref{action} for a metric perturbation $h_{y}^{\ z}=e^{-i \omega u}\phi(r)$.   The equation of motion for $\phi(r)$ works out to be of the same form as that given in~\cite{hmr}:
\be
\frac{f}{r^5}\frac{d}{dr}\left(r^5 f \frac{d\phi}{dr}\right)+ \frac{(1-f)\gamma^2\omega^2}{r^4}\phi(r)=0\,,
\ee
with
\be
f=\left( 1 - \frac{r_0^2}{r^2} \right)
		\left( 1+ \frac{r_0^2}{r^2} - \frac{Q^2}{r_0^2 r^4} \right)
\,.
\ee
Using this, it is straightforward to check that
\be
\frac{\eta}{s}=\frac{1}{4\pi}\,,
\ee
as expected. The calculation in the extremal case proceeds similarly except that the near horizon boundary condition gets replaced by
\be
\phi(\omega, r)=\phi_0 e^{-\frac{i \omega \gamma}{12 (r-r_0)}}(1+\chi_1(r))\,,
\ee
with
\be
\chi_1(r)=-\frac{i \gamma}{36 r_0(r+r_0)}\left(-3 r_0+2(r+r_0)\log \frac{r^2-r_0^2}{r^2+2r_0^2}\right)\,.
\ee
Repeating the calculation for the two-point function (following the steps in \cite{hmr} for example) we again get $\eta/s=1/4\pi$ at $T=0$. This is the first explicit calculation for $\eta/s$ at $T=0$ for any system and matches with the extrapolation intuition gained in \cite{mps09} for the $\gamma=0$ case.

\section{Soliton}\label{s:soliton}

Another interesting solution with Schr\"odinger asymptotics that can be obtained in a straightforward manner is the TsT-transformed soliton.%
\footnote{We are very grateful to Rob Myers for discussions about the topic of this section, and in particular for emphasizing that the soliton phase only exists for certain values of the parameters.} Such solution gives rise to the possibility of getting phase transitions in a planar setting. The soliton is obtained by a double analytic continuation starting from our TsT-transformed background~\eqref{TsT}. Specifically:
\begin{equation}
	x\rightarrow i \hat t\,, \qquad t \rightarrow i \hat x\,, \qquad \gamma \rightarrow i\hat\gamma\,,\qquad Q \rightarrow i\hat{Q}\,,
\end{equation}
will lead to a solution to the equations of motion with Schr\"odinger asymptotics. The soliton can also be obtained by performing the usual TsT transformation on the AdS soliton. The coordinate $\hat x$ has to be periodically identified for regularity. In order to avoid confusion, we will relabel $r_0$ by $r_s$, and in the following we will use the subscript $s$ for quantities related to the soliton and the subscript $b$ for quantities related to the black hole.

Let us study the $(\hat t, \hat x)$ part of the string frame soliton metric:
\begin{equation}
	ds^2=\frac{r^2}{l^2} \hat\M ( f d\hat x^2-d\hat t^2-\hat \gamma^2 r^2 f(d\hat t+d\hat x)^2)\,.
\end{equation}
We want to consider the regularity condition at $r = r_s$. Near $r=r_s$, the metric reads:
\begin{equation}
	ds^2=-\frac{r_s^2}{l^2} (1-\hat\gamma^2 r_s^2)^{-1}
		\left(d\hat t^2+2 \frac{2 r_s^6+\hat{Q}^2}{r_s^7}
		(d\hat x^2-(d\hat t+d\hat x)^2 r_s^2 \hat \gamma^2)(r-r_s)\right)\,,
\end{equation}
and the $O(r-r_s)$ term can be written as
\begin{equation}\label{rrs}
	2(r-r_s)\frac{2r_s^6+\hat{Q}^2}{r_s^7}
		\left[ \left(d\hat x-\frac{r_s^2 \hat \gamma^2 }{1-r_s^2 \hat\gamma^2} d\hat t \right)^2
		+ r_s^2\hat\gamma^2 d\hat t^2
		\left(\frac{r_s^2 \hat\gamma^2}{1-r_s^2\hat \gamma^2}-1 \right) \right] \,.
\end{equation}
Regularity will demand a specific periodicity of the coordinates appearing in the term in square brackets in~\eqref{rrs}. However, asymptotically $\hat v=\frac{1}{2\hat\gamma l}(\hat t-\hat x)$ is already periodic! The compatibility of the two conditions then force us to set:
\begin{equation}\label{rel}
	2 r_s^2 \hat \gamma^2 = 1\,.
\end{equation}
In other words, it only makes sense to consider the soliton for special values of $r_s$ (or $\hat \gamma$). The relation~\eqref{rel} is not modified for $T=0$, which can be seen by expanding to $O((r-r_s)^2)$. We emphasize that this feature is only present in non-relativistic systems. This leads to a crucial difference between the relativistic and non-relativistic case. $\Delta I=I_{\text{sol}} - I_{\text{bh}}$ in the relativistic case ($\gamma\to 0$ or $\mu_1\to \infty$) is a function of $ \mu_2$, $T$ and $r_s$. On the CFT side, $r_s$ controls the radius of the circle on which we are putting the theory and can take any value. In the non-relativistic solutions, \eqref{rel} imposes a relation between $r_s$ and $\mu_1$ which makes $\Delta I$ a function of $\mu_2, T, \mu_1$ with the radius of the circle fixed.

In the usual relativistic setting, corresponding to putting the CFT on a circle with antiperiodic boundary conditions for fermions, the soliton corresponds to the ground state at low temperatures~\cite{sol2}. What happens in the non-relativistic case? As a first step, let us consider the $Q=\hat Q=0$ case that was previously investigated in~\cite{Mann:2009yx}. One can compute the difference between the Euclidean actions of the soliton and the black hole, properly matching the periodicities of the coordinates and putting $\gamma=\hat \gamma$, obtaining:
\begin{equation}
	I_{\text{sol}} - I_{\text{bh}} = - \frac{V \pi^2 \gamma^2 l^6}{2 \kappa^2}
		\left( \frac{r_s^3}{r_b}-\frac{r_b^3}{r_s} \right)
\end{equation}
(we have renamed $r_0=r_b$ for the black hole). There are some crucial differences between our result and that found in~\cite{Mann:2009yx}. Firstly, regularity fixes the periodicity of $\hat x$ to be $\pi l^2/r_s$. This in turn is given in terms of $\hgamma$ in~\eqref{rel}. The black hole $x$ periodicity $\Delta x_b$ had to be matched with the periodicity $\Delta x_s$ of $\hat x$. This means that only for a special choice of $\Delta x_b$ can there be the issue of a phase transition. For general $\Delta x_b$ the boundary conditions will not match and there will be no such phase transition.

Bearing in mind the subtlety due to~\eqref{rel} as discussed above, let us now consider the phase structure when the charged black hole, $Q \neq 0$, is involved. The soliton solution we consider is the one obtained by the double analytic continuation of the \emph{uncharged} TsT-transformed black hole, so $\hat Q =0$, and in order to match the chemical potentials we turn on a constant $A_{\hat t}$.%
\footnote{We could also have considered the double analytic continuation of the charged black hole, $\hat Q \neq 0$, in which case, in order to have the same asymptotics, we should also have turned on a constant $A_x$ in the black hole.}
Our interest is in the phase transition between the black hole at zero temperature and the soliton solution we have just described. The existence of this possibility is due to the presence of the additional tunable chemical potential $\mu_2$ related to the R-charge $Q$, and has not been previously explored even in the relativistic case. The Euclidean action for the soliton turns out to be:
\begin{equation}
	I_{\text{sol}} = -\frac{\Delta \tau_s \Delta x_s  V r_s^4}{2 \kappa^2}
		=  -\frac{\Delta \tau_s \Delta x_s  V}{8 \kappa^2 \hgamma^4}
\end{equation}
where we have used the relation~\eqref{rel}.

For the zero temperature charged black hole, we start from the relations~\eqref{inverse} and take the $T\to0$ limit, getting:
\begin{equation}
	r_0 \xrightarrow{T\to 0} \frac{l^{3/2} \mu_2}{\sqrt{\mu_1}}\,,\qquad
	\frac{Q}{r_0^3} \xrightarrow{T\to 0} \sqrt{2}\,.
\end{equation}
Using these relations, the Euclidean action for the charged TsT-transformed black hole at zero temperature turns out to be:
\begin{equation}
	I_{\text{bh,}\, T=0} = -\frac{\Delta \tau_b \Delta x_b  V\, 3\, l^6 \mu_2^4}{2 \kappa^2 \mu_1^2}\,,
\end{equation}
where we notice that the $\tau$ periodicity goes to infinity in the extremal limit. In order to compute the difference of the actions, we need to match the periodicities and to put $\gamma = \hgamma$ so that there is only one chemical potential $\mu_1$, which is equal for the soliton and the black hole. Furthermore, the black hole $x$ periodicity has to be matched with that of $\hat x$. As discussed above this restricts our discussion to special choices of the circle periodicity. The final result is:
\begin{equation}
	\Delta I=\left( I_{\text{sol}} - I_{\text{bh}} \right)\rvert_{T=0}
		= -\frac{\Delta \tau \Delta x  V\, l^6}{2 \kappa^2 \mu_1^2}
		\left(\mu_1^4 - 3 \mu_2^4\right)\,.
\end{equation}
Thus,  we identify the following phases:
\begin{equation}
\begin{aligned}
	\mu_1 &> 3^{1/4} \mu_2 & & \text{soliton}\,,\\
	\mu_1 &< 3^{1/4} \mu_2 & & \text{black hole}\,.
\end{aligned}
\end{equation}
In the absence of $\mu_2$, the soliton would have been the preferred phase. The novelty here is that when $\mu_2$ becomes sufficiently large, the black hole phase begins to dominate. Thus we have a confinement-deconfinement phase transition at zero temperature. This phase transition is also expected in the relativistic case with a function of $r_s$ replacing $\mu_1$.

\section{Conclusions}\label{s:concl}

In this paper we presented an example of non-relativistic metrics admitting an extremal limit. One of the motivations for studying this solution was the presence of a double zero at the horizon which seems to be relevant for studying Fermi surfaces along the lines of~\cite{liuetal}. It will be interesting to study in what sense the non-relativistic fermion correlations differ from their relativistic counterparts.

We also studied the near horizon limit of the non-relativistic charged black hole background, and found two configurations of fields that solve the equations of motion in the neighborhood of the horizon. It will be of relevance to study the validity of the two configurations to describe different sectors of the dual field theory and in particular to understand better, on the dual side, in which sense the non-relativistic CFT reduces to a non-commutative dipole theory in the specific decoupled AdS$_2$ $\times$ R$^3$ $\times$ $\tilde S^5$ subsector described by~\eqref{scale}. It is tempting to conjecture that ~\eqref{TsTnh} will describe a decoupled subsector of the CFT as well which will include higher energy excitations. More generally, it will be desirable to extend the analysis in \cite{hr2} to the context of non-relativistic extremal black holes, understanding the minimum set of conditions needed for a near horizon solution to exist in the sense presented in this paper. Does the double zero at the horizon guarantee that this will happen?

We considered the TsT-transformed soliton and found that there is a novel phase transition at $T=0$ where, for sufficiently large chemical potential associated with the R-charge, the black hole is the preferred phase. We did not consider the full phase space which includes the ``charged" Schr\"odinger soliton with the extra parameter $\hat Q$. In this case the planar black hole which competes in the path integral will have a constant $A_x$ turned on. On the CFT side, this will have the interpretation of turning on Wilson loops. We leave the exploration of this avenue as future work.

An important consequence of the TsT transformation and of imposing the periodicity of $v$ is that the radius $r_s$ of the Schr\"odinger soliton takes only special values as dictated by~\eqref{rel}. Does it make sense to lift this restriction and ask if there are more general solutions which may in fact dominate in the path integral? In order for this to happen, the relevant geometry should close off before $r=r_s$. It will be very interesting to investigate this possibility.

The R-charged planar AdS black hole has recently featured in another interesting context. As is well known the famous KSS bound conjecture states that the ratio of the shear viscosity to entropy density exceeds $1/4\pi$ \cite{kss}. This has been proven to be true in string theory examples with adjoint matter \cite{bls} in the large $N$, large 't Hooft coupling limit. However, recently it has become clear that in models with fundamental matter \cite{kp} this bound is violated at $O(1/N)$. It was shown in \cite{mps09, cremonini} that in the presence of an R-charge chemical potential, $\ie$ by considering higher derivative corrections to the R-charged planar AdS black hole, this violation is only enhanced. It is tempting to conjecture that this violation will persist even for the non-relativistic examples considered in this paper. Higher derivative corrections in this case will be more nontrivial due to the presence of background fluxes.%
\footnote{An attempt was made in this context for $Q=0$ in~\cite{ghodsi} with the $C^4$ term.}
However, it was found in \cite{mps09} that the relevant higher derivative terms necessarily involve a curvature tensor. Hence it is plausible that writing down a general class of such terms one can extend this analysis to the non-relativistic examples.

\acknowledgments
We are grateful to  Sujay Ashok, Francesco Bigazzi, Alex Buchel, Aldo Cotrone, Sumit Das, Eleonora Dell'Aquila and Malcolm Perry for useful discussions. Special thanks to Allan Adams for discussions on the near horizon limit and to Rob Myers for many discussions on the soliton and for comments on the manuscript. AS thanks Ajay Singh and Jorge Escobedo for collaboration in the initial stages of this project. EI is grateful to the Perimeter Institute for kind hospitality and financial support of his stay, in which this work was initiated. Research at Perimeter Institute is supported by the Government of Canada through Industry Canada and by the Province of Ontario through the Ministry of Research \& Innovation. The work of EI is supported by the Belgian Fonds de la Recherche Fondamentale Collective (grant 2.4655.07). It is also supported in part by the Belgian Institut Interuniversitaire des Sciences Nucl\'eaires (grant 4.4505.86), the Interuniversity Attraction Poles Programme (Belgian Science Policy) and the European Commission FP6 programme MRTN-CT-2004-005104 (in association with V. U. Brussels).

\appendix

\section{Conventions and TsT formulae}\label{s:conventions}

In this appendix we summarize our conventions and review the general formulae for the TsT transformation. The type IIB supergravity lagrangian is written in the Einstein frame as:
\begin{equation}\label{IIBaction}
\begin{split}
	I_{\text{IIB}} &= \frac{1}{2\kappa^2} \bigg[ \int d^{10}x \sqrt{-g}\ R
		- \frac{1}{2} \int \Big( d\Phi \wedge \hd{} d\Phi
		+ e^{-\Phi}  H \wedge \hd{}H \\
		&\qquad\qquad
		+ e^{2 \Phi}  F_1 \wedge \hd{}F_1
		+ e^{\Phi} \mathcal{F}_3 \wedge \hd{}\mathcal{F}_3
		+ \frac{1}{2} \mathcal{F}_5 \wedge \hd{}\mathcal{F}_5
		- C_4\wedge H \wedge F_3 \Big) \bigg] \,,
\end{split}
\end{equation}
where $\kappa = 8 \pi^{7/2} \gs \ls^4$, $H=dB$, $F_p = dC_{p-1}$ and where the modified field strengths $\mathcal{F}_p$ are defined as:
\begin{equation}\label{modifiedF}
	\mathcal{F}_p = F_p + H \wedge C_{p-3}\,.
\end{equation}
The self-duality of the five-form, $\mathcal{F}_5 = \hd{} \mathcal{F}_5$, is not taken into account by the type IIB action and has to be imposed on-shell. The equations of motion that descend from the action~\eqref{IIBaction} are:
\begin{equation}
\begin{gathered}
	d\, \hd{} d \Phi + \tfrac{1}{2} e^{-\Phi} H \wedge \hd{} H
		- e^{2\Phi} F_1 \wedge \hd{} F_1
		- \tfrac{1}{2} e^{\Phi} \mathcal{F}_3 \wedge \hd{} \mathcal{F}_3 =0\,,\\
	d \left[ e^{-\Phi}\, \hd{} H \right] + e^{\Phi} F_1 \wedge \hd{} \mathcal{F}_3
		- \mathcal{F}_5 \wedge \mathcal{F}_3 =0\,,\\
	d \left[ e^{2 \Phi}\, \hd{} F_1 \right] - e^{\Phi} H \wedge \hd{} \mathcal{F}_3 =0\,,\quad
	d \left[ e^{\Phi}\, \hd{} \mathcal{F}_3 \right] + \mathcal{F}_5 \wedge H =0\,,\quad
	d\, \hd{} \mathcal{F}_5 + H \wedge \mathcal{F}_3 =0\,,
\end{gathered}
\end{equation}
and
\begin{equation}
\begin{split}
	R_{MN} &- \tfrac{1}{2} g_{MN} R =
		\tfrac{1}{2} \left( \partial_M \Phi \partial_N \Phi
		- \tfrac{1}{2} G_{MN} ( \partial \Phi )^2 \right)
		+ \tfrac{1}{12} e^{-\Phi} \left( 3 H_{MPQ} H_{N}^{\phantom{N}PQ}
		- \tfrac{1}{2} G_{MN} H^2 \right)\\
		&+ \tfrac{1}{12} e^{\Phi} \left( 3 \mathcal{F}_{(3)\, MPQ}
		 \mathcal{F}_{(3)\, N}^{\phantom{(3)\, N}PQ}
		- \tfrac{1}{2} G_{MN}  {\mathcal{F}_{(3)}}^2 \right)
		+ \tfrac{1}{2} e^{2\Phi} \left( F_{(1)\, M} F_{(1)\, N}
		- \tfrac{1}{2} G_{MN} {F_{(1)}}^2 \right)\\
		&+ \tfrac{1}{96} \mathcal{F}_{(5)\, MPQRS}
		 \mathcal{F}_{(5)\, N}^{\phantom{(5)\, N}PQRS}\,.
\end{split}
\end{equation}

In the main text, we use Einstein frame expressions to check equations of motion and study properties of the solutions such as the thermodynamics. On the other hand, we usually present the solutions in the string frame, where the expressions, as well as the TsT transformation rules, are more easily written. We then recall the conversion of the metric between Einstein and string frame:
\begin{equation}
	ds^2_{\ \text{(string)}} = e^{\Phi / 2}\, ds^2_{\ \text{(Einstein)}}\,.
\end{equation}

Finally, we review the general formulae for the TsT transformation~\cite{Lunin:2005jy} of a type IIB supergravity background, in the explicit form presented in~\cite{ei}. Starting with a solution in the string frame (whose fields we denote with a zero superscript), define $e_{\mu\nu} = G_{\mu\nu}^{(0)} + B_{\mu\nu}^{(0)}$. Identify two $U(1)$ isometries $\varphi^\alpha$ of the solution and perform a T-duality along $\varphi^1$. Next, in the T-dual background, shift $\varphi^2$ as $\varphi^2 \to \varphi^2 + \gamma \varphi^1$, where $\gamma$ is a real parameter, and finally perform another T-duality along $\varphi^1$. The string frame NS-NS fields $E_{\mu\nu} = G_{\mu\nu} + B_{\mu\nu}$ and $\Phi$ of the resulting solution are given by:
\begin{equation}\label{TsT_NSNS}
\begin{split}
	E_{\mu\nu} &= \M \left\{ e_{\mu\nu}
		- \gamma \left[ \det \begin{pmatrix}
			e_{12} & e_{1 \nu}\\
			e_{\mu 2} & e_{\mu \nu}
			\end{pmatrix}
			- \det \begin{pmatrix}
			e_{21} & e_{2 \nu}\\
			e_{\mu 1} & e_{\mu \nu}
			\end{pmatrix} \right]
		+ \gamma^2 \det \begin{pmatrix}
			e_{11} & e_{12} & e_{1\nu}\\
			e_{21} & e_{22} & e_{2\nu}\\
			e_{\mu 1} & e_{\mu 2} & e_{\mu\nu}
			\end{pmatrix} \right\} \,,\\
	e^{2 \Phi} &= \M e^{2 \Phi^{(0)}}\,,
\end{split}
\end{equation}
where
\begin{equation}
	\M = \left\{ 1 - \gamma \left(e_{12} - e_{21} \right) + \gamma^2 \det \begin{pmatrix}
			e_{11} & e_{12}\\
			e_{21} & e_{22}
			\end{pmatrix} \right\}^{-1}\,.
\end{equation}
In the R-R sector, the new modified field strengths $\mathcal{F}_p = F_p + H \wedge C_{p-3}$ are given by:
\begin{equation}\label{TsT_RR}
	\sum_q \mathcal{F}_q \wedge e^{B}
		= \sum_q \mathcal{F}^{(0)}_q \wedge e^{B^{(0)}}
		+ \gamma\, \iota_{\varphi^1} \iota_{\varphi^2}
		\left[ \sum_q \mathcal{F}^{(0)}_q \wedge e^{B^{(0)}} \right]\,,
\end{equation}
The interior product $\iota$ acts on a $p$-form $\omega_p$ giving a $(p-1)$-form with components:
\begin{equation}\label{omegay}
	(\iota_{y}\, \omega_{p})_{\alpha_1 \dotsm \alpha_{p-1}} = (\omega_p)_{y\, \alpha_1 \dotsm \alpha_{p-1}}\,.
\end{equation}
With a suitable gauge choice, \eqref{TsT_RR} can be recast in terms of R-R potentials:
\begin{equation}\label{RRpotTsT}
	\sum_q C_q \wedge e^{B}
		= \sum_q C^{(0)}_q \wedge e^{B^{(0)}}
		+ \gamma\, \iota_{\varphi^1} \iota_{\varphi^2}
		 \left[ \sum_q C^{(0)}_q \wedge e^{B^{(0)}} \right]\,.
\end{equation}
Formulae~\eqref{TsT_RR} and~\eqref{RRpotTsT} are to be understood as formal expressions, valid order by order in the degree of the differential forms.

\section{Near horizon geometry as a solution to the equations of motion}\label{s:nheoms}

In this appendix we will show in more detail why the extremal limit allows for a truncated set of fields which solve the equations of motion in the near horizon limit. For simplicity we will restrict ourselves to the AdS case. Let us consider the action
\be
 I=\frac{1}{2\lp^3}\int d^5x
\sqrt{-g}\left[\,\frac{12}{l^2} + R
-\frac{1}{4}F^2\right]\,.
 \ee
The equations of motion read
\be
R_{ab}-\half R g_{ab}\ =\ \half F_{ac}F_b{}^{c}
-\frac{1}{8} F^2 g_{ab}+\frac{6}{l^2} g_{ab} \,,
\ee
and
\be
\nabla_\mu F^{\mu \nu}=0\,.
\ee
The AdS charged planar black hole considered in this paper is a solution to the equations of motion. Consider the ansatz
\begin{equation}
\begin{split}
ds^2&=-f(u) dt^2+ g(u) du^2+ \frac{r_0^2}{l^2}(dx^2+dy^2+dz^2)\,,\\
A_t&= a(u)\,.
\end{split}
\end{equation}
Here we are anticipating a horizon at $r=r_0$ and are using the coordinates $u=r-r_0$. We will further demand that $a(0)=0$, $a'(0)={\rm constant}$.
The equations of motion lead to the following constraints
\be\label{cons2}
24 g f= a'^2 l^2\,,\qquad f'' a'-f' a''-(a')^3=0\,.
\ee
Now consider a Taylor expansion of $f$ and $a$
\be
f(u)=\sum_{n=1}^{N} f_n u^n\,,\qquad a(u)=\sum_{m=1}^{M} a_m u^m\,,
\ee
where $N$ and $M$ are finite. We want the second of \eqref{cons2} to be satisfied. At $O(u^0)$ we get
\be
2 f_2 a_1= 2 f_1 a_2 +a_1^3\,.
\ee
Using the exact known forms of $f$ and $a$ we find that this can only be satisfied for the extremal case. We have thus proven that only the extremal solution allows for a near horizon configuration of fields that solves the equations of motion. We should however point out that this proof may be coordinate dependent and it is possible that a cleverer choice of coordinates will allow for other solutions.

\section{Isometries of the near horizon geometry}\label{s:isometries}

In the main text, we have presented two geometries that describe the near horizon region of the black hole~\eqref{TsT} in the extremal limit. While the background~\eqref{scale} contains an AdS$_2$ factor, the ``non-relativistic near horizon limit''~\eqref{TsTnh} does not, so we devote this appendix to a more careful study of its symmetries. The metric involving the $x,t,\rho$ part of~\eqref{TsTnh} reads:
\be
ds^2 =\frac{r_0^2}{l^2} \M_0
		\left( - \frac{12 \rho^2}{r_0^2} dt^2 + dx^2 - 12\gamma^2 {\rho^2}   (dt + dx)^2 \right) + \frac{l^2}{12}\frac{d\rho^2}{\rho^2}\,.
\ee
We are looking for transformations
\be
t\rightarrow t+\xi_t(t,\rho,x)\,,\qquad x\rightarrow x+\xi_x(t,\rho,x)\,,\qquad \rho\rightarrow \rho+\xi_\rho(t,\rho,x)\,,
\ee
which leave the metric invariant. This leads to six equations, corresponding to the coefficients of $d\rho^2$, $dt^2$, $dx^2$, $d\rho dt$, $d\rho dx$, $dt dx$. The equation corresponding to the coefficient of $d\rho^2$ reads
\be
\rho \partial_\rho \xi_\rho-\xi_\rho=0\,,
\ee
which leads to $\xi_\rho=\rho R(t,x)$. Plugging this into the remaining equations and looking at the coefficients of $dt d\rho$, $dx d\rho$ leads to setting
\begin{equation}
\begin{split}
\xi_t&= T(t,x)+ \frac{l^4}{288\rho^2} \left(24 \rho^2 \gamma^2 \log\rho \partial_x R-(1+24\rho^2 \gamma^2 \log\rho)\partial_t R\right)\,,\\
\xi_x &= X(t,x) -\frac{l^4 \log\rho}{12 r_0^2} \left( (1+r_0^2\gamma^2)\partial_x R -r_0^2\gamma^2 \partial_t R\right)\,.
\end{split}
\end{equation}
Plugging these into the coefficient of $dt^2$ leads to
\begin{multline}
- 288\rho^2 (1+r_0^2 \gamma^2) R \\ -\M_0^{-1} \left(288\rho^2 (1+r_0^2\gamma^2) \partial_t T+288 \rho^2 r_0^2 \gamma^2 \partial_t X-l^4(1+r_0^2\gamma^2+24\rho^2 \gamma^2 \log \rho)\partial_t^2 R\right)=0\,.
\end{multline}
Now noting that $R$, $X$, $T$ are only functions or $(t,x)$ we are led to setting $R=0$. Solving for $X$, $T$ we get
\be
T=-\frac{r_0^2\gamma^2 X}{1+r_0^2 \gamma^2} + \tau(x)\,.
\ee
Plugging this into the equations arising from $dx^2$ and $dt dx$ and invoking similar arguments leads to setting $\tau$ and $X$ to be constants. Hence we are led to only the translational isometries being preserved. It can be checked easily that the flux configuration also respects these isometries.

\end{document}